\documentclass[aps,prd,onecolumn,superscriptaddress,nofootinbib,eqsecnumm]{revtex4}
\usepackage[pass,letterpaper]{geometry}
\usepackage{graphicx,color}
\usepackage{latexsym,amsmath,amsfonts,amssymb,booktabs}
\usepackage[font=small]{caption}
\usepackage[pdfencoding=auto]{hyperref}
\usepackage{slashed,upgreek,amscd,cancel,tensor,color}

\newcommand{\be}{\begin{equation}}
\newcommand{\ee}{\end{equation}}
\newcommand{\beq}{\begin{equation}}
\newcommand{\eeq}{\end{equation}}
\newcommand{\bea}{\begin{eqnarray}}
\newcommand{\eea}{\end{eqnarray}}

\newcommand{\baryon}{{\rm b}}

\newcommand{\obh}{$\Omega_{\baryon}h^2$}

\newcommand{\npg}{{$^1$H(n,$\gamma)^2$H}}
\newcommand{\ddn}{{D(d,n)$^3$He}}
\newcommand{\ddp}{{D(d,p)$^3$H}}
\newcommand{\dpg}{{D(p,$\gamma)^3$He}}

\newcommand{\sfac}{$S$-factor}

\newcommand\ees{\end{eqnarray}}
\newcommand\bees{\begin{eqnarray}}

\begin{document}
\title{Resolving conclusions about the early Universe requires accurate nuclear measurements}

\author{
  Cyril Pitrou,$^{1}$\thanks{pitrou@iap.fr}
  Alain Coc,$^{2}$
  Jean-Philippe Uzan,$^{1}$
  Elisabeth Vangioni$^{1}$
\vskip0.25cm
$^{1}$Institut d'Astrophysique de Paris, CNRS UMR 7095, 98 bis Bd Arago, 75014 Paris, France\\
Sorbonne Universit\'e, Institut Lagrange de Paris, 98 bis Bd Arago, 75014 Paris, France\\
$^{2}$IJCLab, CNRS IN2P3, Universit\'e  Paris-Saclay, B\^atiment 104, F-91405 Orsay  Campus France
}


\begin{abstract}
Nuclear physics experiments give reaction rates that, via modelling and comparison with primordial abundances, constrain cosmological parameters. The error bars of a key reaction, \dpg, were tightened in 2020, bringing to light discrepancies between different analyses and calling for more accurate measurements of other reactions.
\end{abstract}

\maketitle

One of the key parameters of the standard cosmological model is its density of baryons, a quantity that can be inferred by analysing~\cite{Planck2018} the cosmic microwave background (CMB) or independently by studying Big Bang nucleosynthesis (BBN), the set of reactions that produced atomic nuclei from these baryons in the primordial Universe. The relevant measurements have dramatically improved in the past decades, and been complemented by new observational probes. As a consequence, the error bars on cosmological parameters have been reduced, leading to the era of precision cosmology. However, tensions between different observations remain. One of the key steps in BBN is the \dpg\ reaction --- that is, the fusion of deuterium with a proton resulting in a helium-3 nucleus and a photon, ${\rm d} + p \rightarrow {}^3{\rm He} + \gamma$  --- which partially controls the abundance of deuterium in the first minutes after the Big Bang.  The accuracy of the cross section of the \dpg\ reaction was recently improved~\cite{NatureDPG}, relaunching a debate on the baryon density and divergences between its estimation from BBN and the CMB. 
We wish to highlight the importance for cosmology of refined measurements and statistical analyses of three nuclear rates, \dpg, \ddn\ and \ddp, and draw attention to the reasons of recent diverging conclusions by different teams.

The reanalysis of BBN theoretical predictions after the new \dpg\, measurement~\cite{NatureDPG} has led to two different conclusions. On the one hand, a series of works~\cite{Pisanti:2020efz,Yeh:2020mgl} claim that it confirms the perfect agreement of  BBN with other estimates of the baryon density, \obh, where $h$ is the reduced Hubble constant. On the other hand, our analysis~\cite{Pitrou:2020etk} argues that a new, albeit mild, tension exists between the value this parameter deduced from BBN analysis and the one derived from the CMB and large scale structure data. 

All the reaction rates required for the prediction of the early-Universe abundance of low-mass nuclei are measured in accelerators. Assuming the Copernican principle, the theory describing BBN has only two free parameters, namely  \obh\, and the effective number of relativistic particles $N_{\rm eff}$. Prior to WMAP (the Wilkinson Microwave Anisotropy Probe) in 2003, both were adjustable but \obh\ is now independently determined with high accuracy from the CMB. Hence, one can either fix  \obh\, from CMB and $N_{\rm eff}$ from particle physics (thus making BBN a parameter-free model) or instead constrain the two parameters from BBN (by confronting the predicted abundances to the measured ones) and assess the agreement with other probes. Therefore, the comparison of BBN and CMB at the percent level offers an important test of consistency of the standard cosmological model.

Of the low-mass nuclei produced in the early Universe, not all put strong constraints on \obh. Today, the prediction of helium-4 abundance reaches the percent level and is in full agreement with its observed value.
Given its mild dependence on \obh\ and the accuracy of its measurement, further study of it will not tighten the constraints on baryon density. 
Helium-3 is not very constraining because it is both produced and destroyed in stars so that the evolution of its abundance in time is not very precise.
Lithium-7 exhibits a factor $\sim3$ discrepancy, which is usually discarded quietly, the consensus being that it cannot arise from the nuclear sector. Finally, deuterium is a very fragile isotope that can only be destroyed after BBN. The most recent recommended observed value~\cite{Coo18} is ${\rm D}/{\rm H} = (2.527 \pm 0.030) \times 10^{-5}$ at a redshift $z\sim 2.5-3.1$. Deuterium is thus the most constraining BBN observable because both its observational measurement and its theoretical prediction reach 1\% accuracy. 

Theoretical predictions of abundances depend both on the baryon abundance and the adopted values of nuclear reactions rates. When the baryon abundance is set to its CMB-determined value, if the \dpg\ rate from the NACRE--II database~\cite{NACRE2} is used, an abundance ${\rm D}/{\rm H} = (2.57 \pm 0.13) \times 10^{-5}$ is predicted~\cite{Fie20}, whereas using the new \dpg\ rate~\cite{NatureDPG} (the so-called `LUNA' rate) yields a lower value ${\rm D}/{\rm H} = (2.51 \pm 0.11) \times 10^{-5}$ (Ref.~\cite{Yeh:2020mgl}). Similarly, using the {\tt PArthENoPE} code,
the new  \dpg\ (LUNA) rate leads to ${\rm D}/{\rm H} = (2.52 \pm 0.07) \times 10^{-5}$ in Ref.~\cite{NatureDPG}, and ${\rm D}/{\rm H} = (2.54 \pm 0.07) \times 10^{-5}$ with a re-evaluated LUNA-based rate in Ref.~\cite{Pisanti:2020efz}, whereas our independent analysis using the code PRIMAT~\cite{Pitrou2018PhysRept} and the LUNA rate leads to ${\rm D}/{\rm H} = (2.439 \pm 0.037) \times 10^{-5}$. Consequently, the first two series of works~\cite{Yeh:2020mgl,Pisanti:2020efz} argue that BBN is in agreement with CMB, whereas Ref.~\cite{Pitrou:2020etk} concludes that there is a tension, which can equivalently be considered as a $1.8\sigma$ tension on \obh\ between the BBN and CMB-inferred values.

The production of deuterium mostly depends on 4 nuclear reactions. First, deuterium is produced through \npg, the cross-section of which is obtained from effective field theory computations~\citep{AndoEtAl2006} and is reliable at the 1\%-level.
Then, the three nuclear reactions  \dpg, \ddn\ and \ddp\ are the main sources of nuclear uncertainty for the prediction of the primordial deuterium. Matching the accuracy of observational data requires a percent accuracy on the predictions, which in turn requires reaching a percent level accuracy on the cross sections of these reactions. Achieving such accuracy is a tremendous challenge for nuclear physics, requiring a tight control of systematic errors. Because none of them have resonances, the problem reduces to determining the slowly varying energy-dependent $S$-factors and to the precise determination of their absolute scale. Because all works and codes consider the same physics and the same data, the differences lie in the modelling of this $S$-factor, the selection of datasets, the statistical analysis of uncertainties and on assumptions to fit it to the raw nuclear data. (Existing codes also differ in their treatments of weak interaction rates, which interconvert neutrons and protons, but differences never exceed 0.2\% in the relevant range of temperature and are thus a subdominant source of disagreement.)

Because discrepancies exist between calculations that use the same rate for \dpg, it follows that the source of these discrepancies is the \ddn\ and \ddp\ rates~\cite{Pisanti:2020efz}. For these two reactions, {\tt PRIMAT} relies on the rates based on theoretical, ab initio energy dependences, re-normalized to a selection of experimental data for which estimations of systematic errors are provided. The updated {\tt PArthENoPE} code instead uses a polynomial expansion of the \sfac\ and excludes fewer datasets in fitting its coefficients. Both choices and methods are reasonable and well motivated, and it is a priori impossible with the current data to determine which one is the most appropriate. Hence the disagreement between the adopted nuclear rates can only be reduced with new and accurate nuclear data, for which the effect of these fitting choices will be effectively irrelevant.

In other words, nuclear data are at the crux of the debate. Because differences between BBN codes are attributed to different choices made when modelling the nuclear cross-sections, and not on weak rates, it is important to control their accuracy at least at the percent level and to take into account the latest data. Given the recent LUNA data, the next step is to obtain higher precision measurements and analyses of the  \ddn\ and \ddp\ rates in order to control the accuracy of the different proposed $S$-factors. We also note that the D/H measurement rests heavily on Ref.~\cite{Coo18} and more observations are needed not only to decrease statistical errors but also to eventually reveal subtle systematics. 

This debate on the status of BBN illustrates the interplay between laboratory nuclear physics and astrophysics. It shows that pushing cosmological predictions at the percent level requires more data and a better control of choices that had no significant effect at lower precision. Only then will one be able to conclude if the mild tension we note is the seed of a sharper compatibility issue between CMB or BBN, or if it only reflects the limit of BBN predictions. 


\end{document}